# Possible mechanism of Cooper pairing in HTS cuprates

Avto Tavkhelidze

*Tbilisi State University, Chavchavadze Avenue 13, 0179 Tbilisi, Georgia*

E-mail: *avtotav@gmail.com*

In this study, the possible pairing mechanism based on attraction between electrons from adjacent $CuO_2$ layers is proposed. Initially, each $CuO_2$ layer was found to expand the Fermi sphere owing to ridged geometry. When the two layers are close enough for tunnelling, it becomes energetically advantageous to form correlated quantum states (CQS), reducing the Fermi sphere volume. Cooper pairs, comprising inter-tunnelling electrons, occupy the CQS. The image force is responsible for the electron-electron attraction. Energy exchange between the paired electrons happens through photons. Pair-binding energy and the corresponding effective mass vary in a wide range. At $T>0$, some heavy pairs do not condense. Such pairs are responsible for pseudogap. Light pairs get Bose condensed and are responsible for superconductivity. The proposed mechanism provides possible explanation of two energy gaps and two characteristic temperatures in cuprates. It also provides clarification on the unconventional isotopic effect, Fermi surface pockets, anisotropy of charge transport, and other properties of HTS cuprates. The pseudogap, calculated within the model, is close to the experimental values for the two-layer cuprates, such as YBCO, Bi2212, Tl2212, and Hg1212. It has been shown that the model can be extended to multiple- and single-layer cuprates.

**1 Introduction** Experiments show that in high temperature superconducting (HTS) cuprates, the Cooper pairs [1] are carriers of the superconducting current. However, high $T_c$, low-order parameter, and the unconventional isotopic effect indicate that the phonon mechanism of pairing is not applicable. In HTS cuprates, two separate energy gaps exist [2]. Fermi surface pockets are found in quantum oscillations of hall coefficient [3] and angle-resolving photoemission spectroscopy (ARPES) [4]. $CuO_2$ layers are responsible for superconductivity, and the electrons are concentrated inside them. Reduction of number of $CuO_2$ layers in the ultra-thin films leads to decrease in $T_c$. Furthermore, superconductivity vanishes when less than two layers are left [5], indicating that superconductivity does not exist in a single layer and emerges from some interlayer effect. In this study, the possible pairing mechanism, based on single-electron tunneling between adjacent $CuO_2$ layers, is proposed.

Recently, it was found that the ridged metal films exhibit unconventional properties. Ridges impose additional boundary conditions on the electron wave function and some quantum states become forbidden. Rejected electrons occupy quantum states with higher energies. The Fermi vector, $\mathbf{k}_F$, and Fermi energy, $E_F$, are increased in the ridged geometry [6, 7], which can be termed as Fermi sphere expansion (FSE), for convenience. The pairing mechanism presented in this study is based on the assumption that $CuO_2$ layers, like ridged metal films, exhibit FSE. Cuprate material is divided into $CuO_2$ layers, each containing electron gas modified by FSE. Subsequently, the interaction of the adjacent layers, through single-electron tunneling, has been considered. Cooper pairs exist in correlated quantum states (CQSs), and such states belong to the system of two or more $CuO_2$ layers. In our model pairs do not exist in one particular layer, as in the Lawrence–Doniach model [8] or the electron confinement model [9]. The electron–electron attraction is caused by the image force.

The objectives of this study are to describe a new mechanism of Cooper-pair formation, calculate the pseudogap value on its base, and compare it with the experimental results. First, the mechanism of CQS formation is demonstrated, and subsequently, the mechanism of electron–electron attraction is introduced. Further, the reduction of system energy owing to CQS is calculated. The result is compared with the experimental values of pseudogap for some two-layer cuprates. The possibility of Fermi surface pockets within the model is demonstrated. Finally, the number of unconventional experimental dependences is discussed within the model.

In his study, we use some approximations such as $CuO_2$ layer has ridged geometry, model of nearly free electrons and wide quantum well approximation.

## 2 Mechanism of electron-electron attraction

We regard free electron in confining potential having ridged geometry Fig 1a. Let us name such system as a ridged quantum well (RQW). Fig. 1b shows reference quantum well (QW) having same volume and consequently the same number of free electrons. Periodic ridges on the surface of the RQW layer impose additional boundary conditions on the electron wave function and reduce the quantum state density (DOS). Electrons, rejected from forbidden quantum states, have to occupy the states with higher energy. As a result, Fermi energy in RQW increases from $E_F$ to $E_F^R$.

QS with higher **k**, following Pauli's Exclusion Principle. Consequently, Fermi vector increase and the Fermi sphere expand.

The total energy of the electrons is increased in RQW with respect to QW. The electron gas in RQW is an excited system. If there was some external mechanism to allow back the forbidden QS, then the electrons would occupy them and $E_F^R$ would get decreased (to minimize the energy of the system). We consider electron tunneling to adjacent RQW as a possible mechanism of energy minimization. Fig.2a shows two RQW placed close enough for tunneling and corresponding energy diagram (Fig. 2b). Let us return to our imaginary electron, moving towards the ridged wall. Now, it can tunnel to another RQW. It does not necessarily have to reflect back from the ridged wall, as the tunneling provides

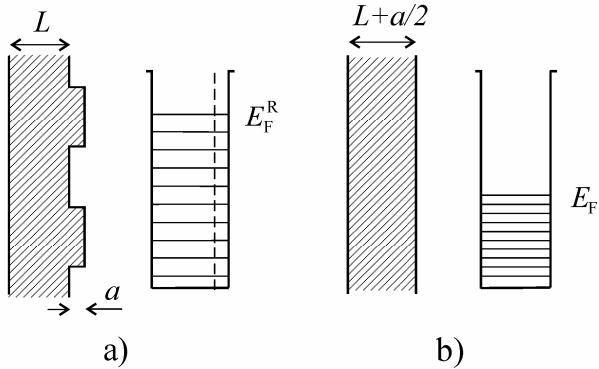
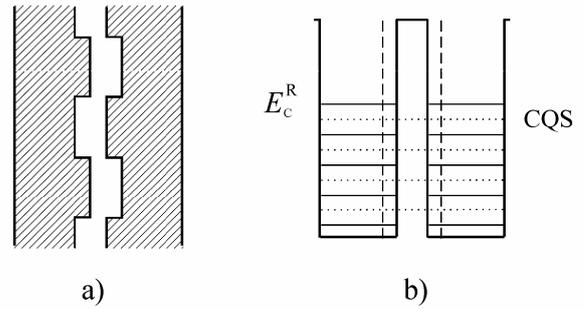

**Figure 1** a) Cross section of RQW and the corresponding quantum energy levels; b) Cross section of reference QW and the corresponding energy levels. Levels are shown as equidistant for simplicity.

**Figure 2** a) Cross section of two RQWs placed close enough for electron tunneling b) Corresponding energy diagram (horizontal lines depict energy levels). Dotted lines depict the correlated quantum states.

For a brief description of the effect, let us consider an electron with energy $E \ll E_F^R$, moving toward the ridged wall (normal to it) as a planar de Broglie wave. Let the value of the electron-wave vector be $\mathbf{k} = \pi/2a$, where $a$ is the depth of the ridge. Subsequently, the two waves, reflected back from the top and the bottom of the ridge, interfere destructively. As a result, the de Broglie wave (and electron) can not reflect back from the ridged wall [6]. Electron can not leave the RQW as it do not has sufficient energy to overcome confining potential. Diffraction of de Broglie wave (reflection under arbitrary angle) is also forbidden, since all the quantum states (QS), which the electron can occupy after diffraction, are already occupied in the Fermi gas. Consequently, all the possible final QS for that imaginary electron are forbidden. Therefore, initial QS also becomes forbidden. All the QS having the wave vector component $\mathbf{k}_x^{(n)} = (\pi/a)(n+1/2)$, where $n$=0, 1, 2… become forbidden, and the DOS gets reduced. If free-electron number is predefined by the requirement of electrical neutrality (e.g., in metal film), some electrons have to occupy one more final QS. More precisely, additional RQW changes the boundary condition for the electron wave function so that non-zero value becomes allowed outside the RQW. Modification of the boundary condition re-establishes the QSs that were forbidden by the ridges. The DOS increases back and the Fermi vector decreases back (Fermi sphere shrinks back).

Adding another RQW reduces the total energy of the electron gas in initial RQW. Closer it is placed, the higher is the probability of tunneling. The probability of tunneling or probability of electron being in the re-established QS increases with the decreasing distance between the RQWs. Consequently, the adjacent RQWs tend to collapse the gap (to reduce the system energy as much as possible). This corresponds to the attractive force. Tunneling occurs in both the directions (because of symmetry) and re-established QS cannot be ascribed to a particular RQW (dotted lines in Fig. 2b). It belongs to the system of two. Re-established QS could be named as correlated quantum state (CQS). The probability of occupation of CQS is equal to the tunneling probability (which is always much less than 1).

Further, we give an additional description of attractive force. QW can be divided into equal parts in two different ways (cross-section is shown in Fig. 3). First, as shown in Fig. 3a, it is divided by plane, resulting in two conventional QWs. Both QWs contain half of the initial number of atoms and free electrons. The Fermi energy of two parts is equal and do not differ from the Fermi energy of the initial QW (within the limit of wide QW approximation). Separation does not change the energy per electron and there is no attraction force between the parts. Subsequently, we can divide the same RQW by ridged plane, as shown in Fig. 3b. Here, the Fermi energy of both parts increases in the process of division, $E_F^R > E_F$. Furthermore, energy per free electron in both the RQWs increases. Now, the parts attract each other to retain the initial unity and reduce the system energy. The only difference between the final states in Fig. 3a and Fig. 3b is the electron-gas energy spectrum, and hence, the attraction force originates from it.

Consider that the electron being in the CQS is tunneling from left to right RQW (Fig. 4). When the electron is inside the barrier, its positive images [10] are created in both the RQWs. The electron image is the mathematical representation of the "transport electrons" redistribution inside the RQW ("transport electrons" are those with energies clse to Fermi energy $E \approx E_F^C \pm K_B T$). As CQS electron passes through the barrier, the right image approaches it and the left image moves away from it, and both the images attract the electron. Let us concentrate on the right image and right RQW. Potentially, the right image can attract one more electron from the right RQW. Thus, the image can potentially serve as a mediator between two electrons (one inside the barrier and another inside the right RWQ) and attract them to each other (like positively charged atom centre in BCS theory). Yet, this is not possible under conventional conditions (target electron from the right RQW is "transport" one). Right image is created by "transport" electrons from the right RQW themselves, and obviously, the image cannot attract one of its own sources. However, in the case of CQS electron, the situation gets principally different. Electrons being in the CQS are those with energies $E \ll E_F^C$ and, therefore, do not participate in image formation. Usually, they do not participate in charge transport (as all adjacent energy levels are fully occupied). Hence, electron being in CQS can be attracted by the image. Thus, the proposed electron–electron attraction mechanism is as follows: CQS electron with wave vector **k**, attracts its right image in the process of tunneling. Right image itself attracts another CQS electron from the right RQW having wave vector, −**k**. As a result, the attraction between two CQS electrons (one being inside the barrier and the other being inside the right RQW) takes place. Since the electron image is only the mathematical representation of the "transport electron" redistribution in space, the real mediator between paired CQS electrons is a collective movement of those "transport electrons". Since electrons are mediators we have photon mechanism of electron-electron attraction.

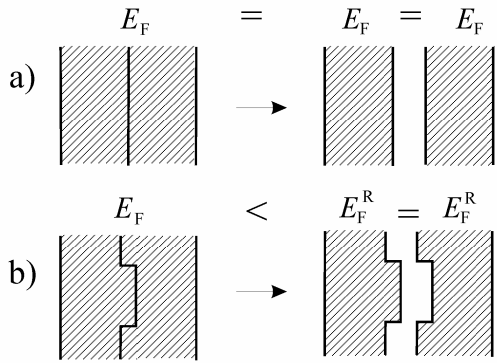

**Figure 3** Two ways of splitting: a) Parts do not attract each other; b) Parts attract each other.

It should be noted that the described electron–electron attraction cannot take place in the system of two conventional QWs. First, it will not work for electrons with $E \ll E_F$, since all QS in both the QWs are already occupied in that energy range (tunneling requires empty QS in the destination QW). Second, it will not work for electrons with energies, $E \approx E_F^C \pm K_B T$ and $E \gg E_F^C$ (both having empty QS around), since these electrons participate in the image formation themselves.

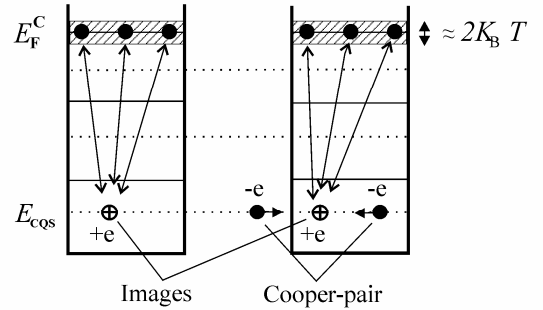

**Figure 4** CQS electron in the process of tunneling between two RQW and image-mediated electron–electron attraction.

Electron being in the CQS can have wave vectors **k** or −**k** (tunneling from left to right or in opposite direction). Taking into account electron spin, there are four possible QS, **k**↑, **k**↓, −**k**↑ and −**k**↓. Utmost, four inter-tunneling pairs, **k**↑ −**k**↑, **k**↑ −**k**↓, **k**↓ −**k**↑, and **k**↓ -**k**↓ could be constructed from them. However, the first and last ones should be excluded, since electrons with the same spin cannot be placed close in real space. The remaining two are Cooper pairs. Therefore, CQS occupied by a maximum of two Cooper pairs, $\mathbf{k}\uparrow -\mathbf{k}\downarrow$ and $\mathbf{k}\downarrow -\mathbf{k}\uparrow$ re-establish to reduce the total energy of the system. The pairs and generally CQS do not remain stationary, since the tunneling probability is low. On the other hand, the density of CQS is high and the product results in stable number of Cooper pairs existing at the same time.

**3 Cooper pairs in cuprates and pseudogap** In cuprates, O and Cu atoms are shifted up and down, relative to the common plane of $CuO_2$ layer. The geometry of the layer is akin to the periodic ridges of RQW. Although the $CuO_2$ layer has no firm boundaries, it is evident that its boundaries are not planar. The boundaries do have some geometry, even in the Hg-based cuprates, where the centers of Cu and O atoms are exactly in the same plane [11]. However, the geometry exists owing to different radii of Cu and O atoms (ions). Hence, in the first approximation, a $CuO_2$ layer (for all cuprates) can be regarded as an RQW-containing electron gas, and the layer has forbidden QS and expanded Fermi sphere. The FSE forces the electron gases in the adjacent layers to reduce their total energy, by means of CQS.

To test the model, we calculate reduction of energy per electron in the system of two $CuO_2$ layers and compare it with measured pseudogap values. Fig. 5 shows two $CuO_2$ layers separated by distance $d$. The density of QS as found in [7] is

$$\rho_{RQW}(E) = \rho_{QW}(E)/G . \qquad (1)$$

Here, $\rho_{RQW}(E)$ is the density of QS in RQW, $\rho_{QW}(E)$ is the density of QS in reference QW, $E$ is the electron energy, and $G$ is the geometry factor. Thus, the density of forbidden QS is

$$\rho^-(E) = \rho_{QW}(E) - \rho_{RQW}(E) = \rho_{QW}(E)(1 - G^{-1}) \qquad (2)$$

The probability of electron being in CQS is equal to the tunneling probability. Last can be given as

$$D(E) = \exp\left[-\frac{2d}{\hbar}\sqrt{2m(U-E)}\right] \qquad (3)$$

where, $\hbar$ is the Plank's constant, $m$ is the electron mass, and $U$ is the height of the potential barrier. We assume that the receiving QS is empty, i.e. $1 - D(E) \approx 1$, as $D(E) \ll 1$. "Diving" the electron into CQS leads to energy reduction and binding of adjacent layers. Thus, the layer binding energy density (per unit volume) within the energy interval of $\delta E$ will be

$$\delta E_{bin} = 4\varepsilon_{CQS} D(E) \rho^-(E) \delta E \qquad (4)$$

where $\varepsilon_{CQS} = (U - \varphi - E)$ is the reduction of energy per electron in the process of CQS formation. The factor 4 comes from degeneracy. Each electron that leaves the Fermi level to CQS can be in four possible QS, $\mathbf{k}\uparrow$, $\mathbf{k}\downarrow$, $-\mathbf{k}\uparrow$, and $-\mathbf{k}\downarrow$. Integration of Eq. (4) over the energy range below the Fermi energy gives

$$E_{bin} = 4 \int_{U-\varphi}^{0} (U - \varphi - E) D(E) \rho^-(E) dE \qquad (5)$$

Inserting Eq. (2) in Eq. (5) results in

$$E_{bin} = 4(1 - G^{-1}) \int_{U-\varphi}^{0} (U - \varphi - E) D(E) \rho_{QW}(E) dE \qquad (6)$$

Further, using well known formula for DOS $\rho_{QW}(E) = m\sqrt{2mE}/\pi^2\hbar^3$ we rewrite (6) as

$$E_{bin} = \frac{4\sqrt{2}\,m^{3/2}(1 - G^{-1})}{\pi^2 \hbar^3} \int_{U-\varphi}^{0} (U - \varphi - E) D(E) \sqrt{E}\, dE . \qquad (7)$$

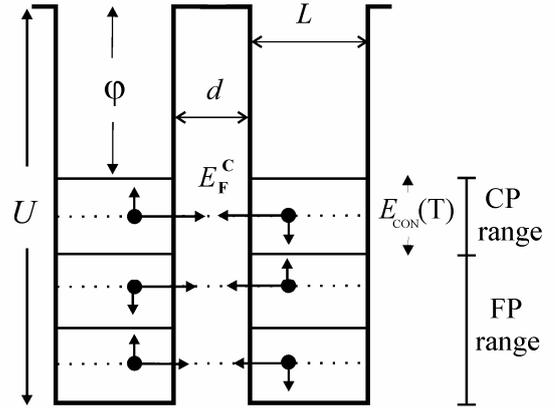

**Figure 5** The CQS (dotted lines) occupied by Cooper pairs.

To get the binding energy per electron, $E_{bin}$ should be divided by the density of the electrons. In hole-doped cuprates, the density of electrons in $CuO_2$ layer is equal to the density of the holes in charge reservoirs, and has the universal value [12] for optimally doped cuprates, $p=1.6\times 10^{21}$ cm$^{-3}$. Consequently, the binding energy per electron in such cuprates is $\varepsilon_{bin} = E_{bin}/p$. Let us calculate the values of $\varepsilon_{bin}$ for some double-layer cuprates and compare it with the measured pseudogap values. Further, let us assumed that $G \gg 1$ and $(1 - G^{-1}) \approx 1$ (strictly, $G$ depends on buckling angle of $CuO_2$ layer, but it can be ignored in first approximation). In Eq. (7), free-electron mass is used, since the effective mass in $c$ direction has no significance inside the single $CuO_2$ layer (there is no periodic lattice potential in $c$ direction inside the layer). The experimental values of interlayer distance for the two-layer cuprates were 3.36 Å for YBCO, 3.35 Å for Bi2212, 3.2 Å for Tl2212 (all three from Ref. 13), and 3.23 Å for Hg1212 (Ref. 11). The given values are the distances between the atom centers. Let the electron cloud radius in tunneling direction be $R_c$. Subsequently, $2R_c$ should be subtracted from the interlayer distance. Thus, we get

$$\varepsilon_{\text{bin}} = \frac{4\sqrt{2}\, m^{3/2}}{\pi^2 \hbar^3 p} \int_{U-\varphi}^{0} (U-\varphi-E)\sqrt{E}\; x$$

$$x \exp\left[-\frac{2(d-2R_c)}{\hbar}\sqrt{2m(U-E)}\right] dE. \quad (8)$$

The following experimental values were inserted in Eq. (8): work function [14], $\varphi = 4\,\text{eV}$; Fermi energy [15, 16], $E_F^C = U - \varphi = 0.3\,\text{eV}$, and interlayer distance, $d$=3.2 Å. The Cu and O atoms have the atomic radii of 1.28 Å and 0.73 Å, respectively, and the ionic radii [17] are 0.87 Å for $Cu^{2+}$ and 1.26 Å for $O^{2-}$. However, it is not clear on which value should be used for effective $R_c$. The natural suggestion is that it should be in the range of 0.73 Å < $R_c$ <1.26 Å. Fig. 6 shows the plot of $\varepsilon_{\text{bin}}$ as the function of $2R_c$, according to Eq. (8), in the abovementioned range of $R_c$.

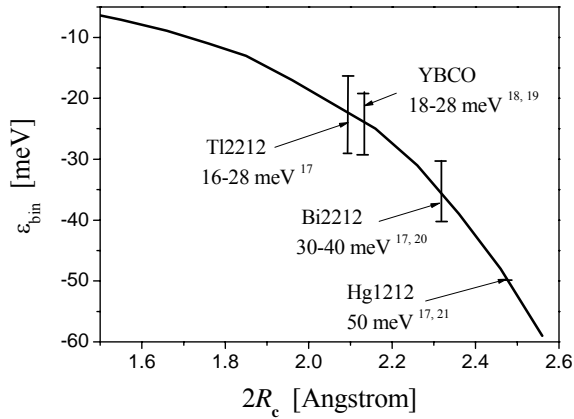

**Figure 6** Solid curve corresponds to binding energy per electron, $\varepsilon_{\text{bin}}$. The experimental values of pseudogap for optimally doped two-layer cuprates are plotted for comparison.

The figure also shows the experimental values of pseudo gaps obtained from the tunneling spectroscopy [17-21]. The values calculated within our model fit the experimental ones in the reasonable range of $R_c$.

It is essential to compare the electron binding energy (in Cooper pair) with the energy reduction (per electron) during CQS formation. The Coulomb attraction between the paired electrons (with positive image +e in the core), gives the binding energy per electron, $(e^2/4\pi\varepsilon_0)(3/4L) = 5\,\text{eV}$, for the layer thickness, $L$= 2 Å. For comparison, the maximum energy reduction during the transition of electron from Fermi level to CQS is only $E_F^C \approx 0.3\,\text{eV}$ in cuprates. Consequently, the image-mediated electron–electron attraction can easily provide the needed energy reduction. It is interesting to note that the latter is not applicable for conventional solids having $E_F \approx 10\,\text{eV}$.

The Fermi energy in cuprates is low and the corresponding de Broglie wavelength $\lambda_F \propto 1/\sqrt{E_F}$ is high. The relatively large $\lambda > \lambda_F = 15$–25 Å allows the electron to participate in the tunneling events at long distances, possibly as far as three $CuO_2$ layers [23, 24]. Therefore, tunneling between three or more layers might contribute significantly. Multi-layer tunneling increases the binding of the layers and allows further reduction in the total energy of the system. This may explain the increase in $T_c$ with the increasing number of similar $CuO_2$ layers per unit cell from 1 to 3.

The described model shows quantitative agreement with the experiment in the case of multi-layer cuprates having $d$=3–4 Å. However, in the case of single-layer cuprates, $d$=6–12 Å and $\varepsilon_{\text{bin}}$ becomes <0.1 meV. Last value is much lesser than the experimental pseudogap values. A possible reason for the high pseudogap value in the single-layer cuprates is the negative-U centers. Residing inside the charge reservoir layers, such centers can serve as resonant tunneling centers and increase the tunneling probability [25], reducing the effective distance between the layers down to 3–4 Å.

**4 Binding energy, Bose condensation and thermal fluctuations** In conventional superconductors, the electron–phonon interaction is responsible for Cooper-pair formation (BCS theory), and the binding energy per electron is of the order of 1 meV. In BCS theory all paired electrons have wave vectors close to Fermi wave vector $\mathbf{k} \approx \mathbf{k}_F$. However, in the presented mechanism, the pair-binding energy is $\Delta_{\text{CQS}}(\mathbf{k}) = E_F^R - E_{\text{CQS}}(\mathbf{k})$ and vary in a wide range of 0–$E_F^C$ ($E_F^C \approx 300$ meV in cuprates) as $|\mathbf{k}|$ varies considerably from pair to pair. The pair-effective mass, $M_{\text{pair}}$, being proportional to $\Delta_{\text{CQS}}(\mathbf{k})$ according to negative-U Hubbard model [26], also vary in a wide range $M_{Pair} \propto [E_F^R - E_{\text{CQS}}(\mathbf{k})]$. Therefore, the pairs have very different effective mass and starting conditions for phase ordering and Bose condensation. For low enough $T$ Cooper pair, having **k** dependent effective mass $M_{pair}(\mathbf{k})$, is either free pair (FP) or condensed pair (CP). This can explain the two energy gaps and the two characteristic temperatures in cuprates. Condensed pairs (CP) result in superconductive gap $\Delta_C$ and free pairs (FP) result in pseudogap $\Delta_P$.

As FPs have more $M_{Pair}$, $\Delta_P > \Delta_C$. The FP and $\Delta_P$ exist below $T^*$, while CP and $\Delta_C$ exist only below $T_C$. Let $E_{CON}(T)$ be the maximum $\Delta_{CQS}(\mathbf{k})$ that allows Bose condensation at a given $T$. The energy interval $E < E_F^C$ can be divided into two regions (Fig. 5), namely the CP region, where $\Delta_{CQS}(\mathbf{k}) < E_{CON}(T)$ and FP region, where $\Delta_{CQS}(\mathbf{k}) > E_{CON}(T)$. At $T=0$, all the pairs are CP, within the whole range $E < E_F^C$. There would be no region of FPs, since at $T=0$ all pairs condense independent of $M_{pair}$. When $T$ gets increased, some pairs leave the condensate owing to high $M_{pair}$, and FPs emerge. With further increase in $T$, the CP region shrinks and disappears at $T=T_C$. However, the FP region remains above $T_C$. Further increase in $T$ reduces the number of FPs owing to thermal fluctuations, and all pairs get destroyed at $T=T^*$.

Both $E_{CON}(T)$ and $T_C$ depends on the phase-ordering mechanism, which is out of scope of this study. Still, one general note can be made. Strong layer binding corresponds to more order and less entropy $S$ of the system. The layer binding energy has not only the CP component, but also the FP component. Consequently, the free pairs influence condensation process indirectly. They increase the layer binding and reduce $S$.

Thermal fluctuations can influence the pairs through the layer geometry. Vibration of Cu and O atoms introduce changes in the geometry ($G$ becomes time dependent). The energy levels move up and down on the energy scale as the RQW geometry follows fluctuations. Since these displacements are not synchronized in the adjacent layers, it redistributes CQS and reduces their probability. At $T>0$, there exists a fluctuation-induced loss of CQS. Therefore, the binding energy of the adjacent layers depends on $T$ and reduces with the increase in $T$. The CQS minimizes the energy of the system, which in terms of atomic shifts signifies that the atoms become more tied to their optimum (for correlated states) position. If there was no adjacent $CuO_2$ layer, then the atom vibration would have added amplitude at the given $T$. Having lesser amplitude of fluctuation shift, at the same $T$, is equivalent to increase in the effective mass. Therefore, the binding of the layers increases the effective masses $M_{Cu}$ and $M_O$ of Cu and O atoms, respectively. This can be the reason for the unconventional dependence of isotopic effect on doping in cuprates. The isotopic effect is minimal for optimal doping [27]. According to our model, the optimal doping corresponds to the highest layer binding and maximum $M_{Cu}$ and $M_O$. The higher the $M_{Cu}$ and $M_O$ are, the less pronounced would be the isotopic effect, as it introduces lesser relative mass change.

## 5 Fermi surface pockets as a consequence of layer geometry
The Fermi surface pockets are found in APRES and Shubnikov–de Haas effect measurements. Our model provides possible explanation for them. The geometry of $CuO_2$ layer strongly modifies the Fermi surface area and shape. The **k** spectrum in the ridged geometry was investigated in [7, 28]. Here, we provide some related details. Fig. 7a shows the RQW and the corresponding **k** spectrum in $\mathbf{k_Y}$, $\mathbf{k_Z}$ plane ($a,b$ plane in cuprates). Electrons having low wave vector component $|\mathbf{k_Y}| < \pi/w$ cannot exist in such geometry. Such de Broglie waves cannot "fit" inside the ridges.

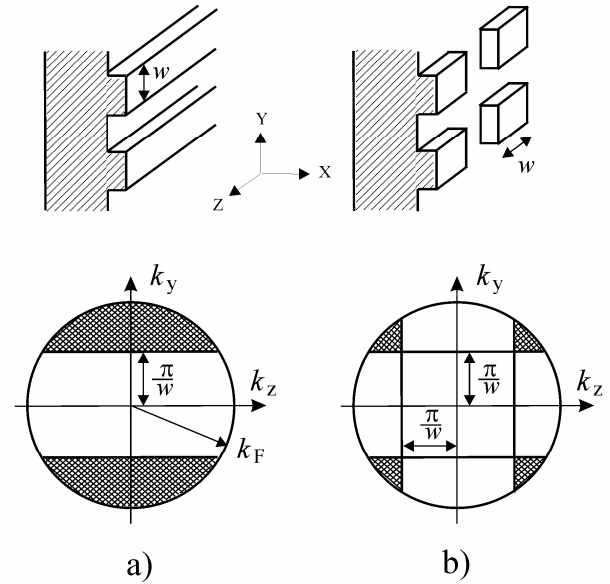

**Figure 7** a). RQW and its $k_y$, $k_z$ spectrum. b) Modified RQW and its $k_y$, $k_z$ spectrum.

Obviously, **k** plane contains external circle of diameter $\mathbf{k_F}$ (maximum possible k at $T=0$). The circle and the two lines $|\mathbf{k_Y}| = \pm\pi/w$ limit the allowed **k** area, shown on the **k** plane as the shaded portion in the bottom of Fig. 7a. Subsequently, the strips are replaced by right-square prisms as shown in Fig. 7b. Here, the $Z$ component of wave vector $\mathbf{k_Z}$ also gets filtered and $|\mathbf{k_Z}| < \pi/w$ also becomes forbidden. As a result, we get four allowed **k** areas, represented by the shaded portion in the bottom of Fig. 7b. The allowed **k** areas in Fig. 7b are akin to Fermi surface pockets observed in cuprates. In our analogy the prisms represent Cu atoms. Obviously, the geometry of electron cloud differs from the prisms and is more like dome. Thus, the shape of Fermi surface pockets should also differ from those shown in Fig. 7(b). Still, the described model provides possible (rough) explanation of the Fermi pockets in cuprates. More precise results (for dome geometry) can be obtained by using special mathematical methods recently developed for Helmholtz spectrum calculation in special geometries [28]. Some useful geometry, including the double-side ridged geometry [29] and the double-side corrugated geometry [30] were investigated.

**6 Comparison with experiments** The model described in this study provides possible explanation of the following cuprate properties:
(1) Low value of order parameter – In our model, the Cooper pairs are concentrated within a few atomic layers (real-space pairing) and consequently, the order parameter is close to 10 Å, which is in agreement with the experiments.
(2) Loss of superconductivity in ultra-thin films – The $T_c$ reduces when the film thickness decreases, and the superconductivity vanishes after less than two $CuO_2$ layers are left [5]. Within our model, at least two $CuO_2$ layers should be present to allow Cooper pairing and superconductivity.
(3) Pseudogap is present in the energy spectrum above $T_c$, and its width does not depend on $T$ – In our model, the pseudogap is formed by free pairs. The pseudogap width does not depend on $T$, since it is consequence of FPs, having pairing energy up to 300 meV, much more than $K_B T^* \approx 10$ meV.
(4) Strong $T^*$ and $T_c$ dependence on oxygen doping and overall deviation from Fermi liquid which scales with pseudogap [31, 32] – In our model, both gaps $\Delta_P$ and $\Delta_c$ depend on the layer geometry, governed by internal pressure or doping. Overall deviation from Fermi liquid can be explained by formation of free pairs (being the reason for pseudogap).
(5) Electronic specific heat, $C_v$ – At $T^*$, the ratio $C_v/T$ starts to reduce with the decrease in $T$ (Ref. 16, 32). It changes behavior at $T^*$, and not at $T_c$ (as for conventional superconductors) - Our model provides possible explanation. When $T$ is decreased, FPs start forming at $T^*$. As FPs thermally decouple from the electron gas, the electronic specific heat decreases.
(6) Scaling relationship between $T_c$ and the buckling angle of the $CuO_2$ planes [33, 34] – Earlier experiments revealed that phonons are not involved in Cooper pair formation. Consequently, it became very difficult to explain the buckling-angle (external pressure) dependence of $T_c$. In our model the buckling angle sets the geometry factor and consequently layer binding energy. This provides possible explanation of observed relationship.
(7) Anisotropy of electrical conductivity $\sigma$ temperature dependence in pseudogap phase – Experiments demonstrate that it is semiconductor-like $(d\sigma/dT)_c < 0$ in $c$ direction and metal-like $(d\sigma/dT)_{ab} > 0$ in $ab$ plane. In our model, the electrons with $\mathbf{k} \approx \mathbf{k}_F$ have large anisotropy in $\mathbf{k}$. Electrons having high $\mathbf{k}_C$ (low $\mathbf{k}_A$ and $\mathbf{k}_B$) participate in the formation of CQSs. Consequently, the electrons with high $\mathbf{k}_C$ are absent in the $\mathbf{k}$ spectrum near $\mathbf{k} \approx \mathbf{k}_F$. Such electrons "dive" from the region $\mathbf{k} \approx \mathbf{k}_F$ into CQS and form Cooper pairs. Empty region $\mathbf{k} \approx \mathbf{k}_F$ results in semiconductor-like behavior of $\sigma(T)$, in $c$ direction. Electrons with $\mathbf{k} \approx \mathbf{k}_F$ and low $\mathbf{k}_C$ (high $\mathbf{k}_A$ and $\mathbf{k}_B$) do not participate in the formation of CQS and, therefore, $\sigma(T)$ dependence in $ab$ plane is metal-like. The scaling of $c$-axis conductivity with pseudogap energy [35] can also be explained within our model.
(8) In experiments, ARPES data is collected only for $\mathbf{k}_A$ and $\mathbf{k}_B$, while $\mathbf{k}_C$ is not measured, as cuprate crystals cleave along the $CuO_2$ plane only. Consequently, the information about $\mathbf{k}_C$ is absent in APRES data [4, 36] (general problem of photo emission spectroscopy). According to our model, the electron pairing introduces changes only in $\mathbf{k}_C$. This provides possible explanation why a large number of precise ARPES data are unable to reveal pairing mechanism so far.
(9) Unconventional isotopic effect – Isotopic effect for both Cu and O atoms scales with the pseudogap, and is minimal at optimal doping [30] - Our model provides possible explanation via binding between $CuO_2$ layers. The large isotopic change of $T^*$ reported in Ref. 37 can also be explained within our model. At high temperatures around $T^*$, the layer binding is weakened by thermal fluctuations. Consequently, the binding-induced increase of $M_O$ is low and the isotopic effect becomes more pronounced.
(10) Recently discovered iron-based HTS material $LaO_{1-x}F_xFeAs$ also has layered structure with conducting AsFe layer [38]. The AsFe layer has the geometry close to the ridged-like $CuO_2$ layer in cuprates.
(11) Fermi surface pockets were observed in ARPES and Shubnikov–de Haas effect measurements – Such pockets might be consequence of the $CuO_2$ layer geometry.

**Conclusions** In this study, the possible mechanism of electron–electron attraction in cuprates, based on image force, is proposed. Electrons tunnel between adjacent $CuO_2$ layers to reduce the energy of the system. Initially, owing to Fermi sphere expansion, the energy of electron gas possesses added value in the individual layers. Electron tunneling allows the formation of CQS, resulting in reduction of system energy, and the Cooper pairs occupy the CQS. At $T>0$, depending on their individual binding energy, some pairs Bose condense, while the others remain free. The condensed pairs are responsible for superconductive gap and free pairs are responsible for pseudogap. The energy reduction per electron, calculated within the model, is in agreement with the experimental values of pseudogap for the two-layer cuprates, such as YBCO, Bi2212, Tl2212, and Hg1212. Presented model provides possible explanation of the two energy gaps and the two character-istic temperatures. It also provides possible explanation to the low-order parameter, Fermi surface pockets, unconventional isotopic effect, conductance anisotropy in pseudogap state, temperature dependence of electronic specific heat, vanishing of superconductivity in ultra-thin films, scaling relationship between buckling angle and $T_c$, and other properties of HTS cuprates.

**Acknowledgments** Author thanks G. Gabadadze and V. Svanidze for their useful discussions. This work was partly sponsored by Borealis Technical Limited.

## References

[1] L. N. Cooper Prys. Rev., **104**, 1189 (1956).

[2] A. Mourachkine, *Room-Temperature Super conductivity*, (Cambridge International Science Publishing, 2004).

[3] S. Chakravarty and H. Kee, arXiv:cond-mat. 0710.0608v3 (2008); N. Doiron-Leyraud et al. Nature **447,** 565 (2007).

[4] A. Damascelli, Z. Shen and Z. Hussain, Rev. Mod. Phys. **75,** 473 (2003).

[5] L. X. You, A. Yurgens, D. Winkler, C. T. Lin, and B. Liang J. Appl. Phys., **98,** 033913.

[6] A. Tavkhelidze et al., J. Vac. Sci. Technol. B, **24,** 1413, (2006)

[7] A. Tavkhelidze, V. Svanidze and I. Noselidze , J. Vac. Sci. Tech. B, **25,** 1270 (2007); A. Tavkhelidze, V. Svanidze, Int. J. Nanoscience v. 7, 333 (2008).

[8] W.E. Lawrence and S. Doniach, *Proceedings of the Twelfth International Conference on Low Temperature Physics, Kyoto 1971*, edited by E.Kanda, (Academic Press of Japan, 1971), p.361.

[9] K. Fukushima and H. Sato, Phys. Rev. B, **46,** 14794 (1992).

[10] D.V. Gepert, J. Appl. Phys., **34**, 490, (1963); Y. Hishinuma, T. H. Geballe, B. Y. Moyzhes, and T. W. Kenny, Appl. Phys. Lett. **78,** 2572 (2001).

[11] T. H. Geballe, G. Koster , arXiv:cond-mat/0604026, (2006).

[12] T. Honma, P. H. Hor, H. H. Hsieh, M. Tanimoto, Phys. Rev. B, **70,** 214517 (2004).

[13] D. M. Ginsberg, *Physical properties of high temperature superconductors* (World Scientific, Singapore, 1994).

[14] K. Nobuyoshi, S. Kazuya, Jp. J. Appl. Phys., **29,** L1635-L1637 (1990).

[15] I. N. Askerzade, Journal of the Korean Physical Society, **45,** 475 (2004) .

[16] E. W. Carlson et al., cond-mat/0206217 v1, (2002).

[17] http://www.chemicool.com/

[18] C. Renner et al. Phys. Rew. Lett., **80**, 3606 (1998).

[19] D. Mazur et al. http://arxiv.org/abs/cond-mat/0611636v1, (2006).

[20] Ø. Fischer et al., Rev. Mod. Phys. **79,** 353 (2007).

[21] Murakami H. et al., Appl. Surf. Sci. **175-176,** 306 (2001).

[22] Wei, J. Y. T. et al. Phys. Rev. B **57,** 3650 (1998).

[23] T. A. Zaleski and T.K. Kopec, Phys. Rev. B, **71,** 014519 (2005).

[24] Theodore H. Geballe , *The never ending search for hight temperature superconductivity*, Chapter III, arXive: cond-mett/0608368 (2006).

[25] V. Oganesyan, S. Kivelson, T. Geballe, and B. Moyzhes, Phys. Rev. B **65** , 172504 (2002).

[26] R. Micnas, J. Ranninger, and S. Robaszkiewicz, Rev. Mod. Phys., **62** , 113 (1990).

[27] Guo-meng Zhao, H Keller and K Conder, *J. Phys.: Condens. Matter* **13,** R569 (2001).

[28] T. Emig, A. Hanke, R. Golestanian, M. Kardar, Phys.Rev.Lett. **87,** 260402 (2001).

[29] R. Büscher and T. Emig, Phys. Rev. Lett. **94** , 133901 (2005).

[30] Robson B. et al., Europhys.Lett. **76** (2006) 822-828.

[31] Di Castro C.; Grilli M.; Caprara S., J. Phys. Chem. Sol., **63,** 2219 (2002) .

[32] J. L. Tallon and J. W. Loram, Physica C, **349** , 53 (2001).

[33] M. Sardar and S. Bhattacharjee, Physica C **349** 89 (2001).

[34] O. Chmaissem *et al.*, Nature **397,** 45 (1999).

[35] Y. H. Su, H. G. Luo, T. Xiang, Phys. Rev. B **73** , 134510 (2006).

[36] J. C. Campuzano et. Al., Phys. Rev. Lett. **83** , 3709 (1999) .

[37] D. R. Temprano, J. Mesot, S. Janssen, K. Conder, and A. Furrer, Prys. Rev. Lett., 84, 1990 (2000).

[38] Y.Nakamura, T. Watanabe, M. Hiranoand H. Hosono, J. Am. Chem. Soc. **130** , 3296 (2008).